


%
%
%

\input iopppt
\jnlstyle
\mediumstyle
\jl{8}

\input epsf


\def\pmb{}


\def\bU{\pmb{U}}
\def\bA{\pmb{A}}
\def\cA{\pmb{\cal A}}
\def\bF{\pmb{\cal F}}
\def\bG{\pmb{\cal G}}
\def\bM{\pmb{\cal M}}
\def\bGt{{\widetilde{\pmb{\cal G}}}}
\def\bLt{{\widetilde{\pmb{\cal L}}}}
\def\htid{{\tilde h}}
\def\htidx{{\tilde h_x}}
\def\htidy{{\tilde h_y}}
\def\hhat{{\hat h}}
\def\hhatx{{\hat h}_x}
\def\hhaty{{\hat h}_y}
\def\ghat{{\hat g}}
\def\Lhat{{\hat{\cal L}}}
\def\rmd{{\rm d}}

\def\kxp{{\kappa_x'}}
\def\kyp{{\kappa_y'}}
\def\rhot{{\tilde\rho}}

\def\Ap{${\rm A}+$}
\def\Am{${\rm A}-$}
\def\Bp{${\rm B}+$}
\def\Bm{${\rm B}-$}

\def\and{\&\ }

\def\half{\hbox{$1\over2$}}

\def\quarter{\hbox{$1\over4$}}
\def\threequarters{\hbox{$3\over4$}}



\def\AMRsemidirectprodsymbol{%
       \mathbin{\vrule height1.10ex depth -0.05ex width0.25pt
                \mkern-2.9mu
                \mathchar"0202}}

\def\AMRsemidirectprod{%
       \mskip-\medmuskip \mkern5mu
       \mathbin{\AMRsemidirectprodsymbol}
       \penalty 900
       \mkern5mu \mskip-\medmuskip}

\hyphenation{co-dimen-sion sol-ution sol-utions pro-pen-sity}


%
%
\def\gapprox{\lower.4ex\hbox{$\;\buildrel >\over{\scriptstyle\sim}\;$}}
\def\lapprox{\lower.4ex\hbox{$\;\buildrel <\over{\scriptstyle\sim}\;$}}


\newcount\AMRreferencecount
\def\ifundefined#1{\expandafter\ifx\csname#1\endcsname\relax}%
\def\addrefnotext#1{\ifundefined{#1}%
  \global\advance\AMRreferencecount by 1%
  \symbol{\the\AMRreferencecount}{#1}
  \expandafter\xdef\csname#1\endcsname{\the\AMRreferencecount}%
\fi}%
\def\addref#1{\addrefnotext{#1}}%

\newwrite\AMRreffile
\immediate\openout\AMRreffile=rs.aux

\def\symbol#1#2{\immediate\write\AMRreffile{#2 #1}}

\newcount \AMRequano%
\global\AMRequano=0
\def \lasteqno{\the\AMRequano}%
\def \AMReqno{\global \advance \AMRequano by 1 {\rm \lasteqno}}%
\def \AMReqlabel#1{\expandafter\xdef\csname #1\endcsname{(\lasteqno)}%
                  \symbol{\lasteqno}{#1}}%
\def \AMRsection#1{\section{#1}}%



\def \lastfigno{Figure~\hbox{\the\figno}}%
\def \nextfigno{{\advance \figno by 1 \lastfigno}}%
\def \AMRfigurelabel#1{\expandafter\xdef\csname #1\endcsname{\lastfigno}%
                      \symbol{\lastfigno}{#1}}%

\newif\ifAMRfigurenaturalsize
\AMRfigurenaturalsizefalse

\newif\ifAMRfigureysize
\AMRfigureysizefalse

\newdimen\AMRfigurewidth
\AMRfigurewidth=\hsize \advance\AMRfigurewidth by -\secindent

\def\AMRfigure #1 #2\par{%
  \topinsert
    \hbox to \hsize{\hfil\begingroup
                \ifAMRfigurenaturalsize
                  \def\epsfsize##1##2{0pt}
                \else
                  \ifAMRfigureysize\else
                    \epsfxsize \AMRfigurewidth
                  \fi
                \fi
                \hbox to \AMRfigurewidth{\hfil\epsffile{#1}\hfil}
                \endgroup}%
    \figure{#2}%
    \AMRfigurenaturalsizefalse
    \AMRfigureysizefalse
    \symbol{\the\figno}{#1}
  \endinsert}%

\newbox\AMRFigurePreboxedBox

\def\AMRfigurepreboxed #1\par{%
  \topinsert
    \unvbox\AMRFigurePreboxedBox%
    \figure{#1}%
  \endinsert}%

\def\lasttabno{Table~\hbox{\the\tabno}}%
\def \nexttabno{{\advance \tabno by 1 \lasttabno}}%
\def \AMRtablelabel#1{\expandafter\xdef\csname #1\endcsname{\lasttabno}}%

\title{Instabilities of periodic orbits with spatio-temporal
       symmetries}[Instabilities of periodic orbits with spatio-temporal
       symmetries]

\author{A M Rucklidge\dag\footnote{\S}{email: A.M.Rucklidge@damtp.cam.ac.uk}
 and M Silber\ddag
 \footnote{\P}{email: silber@nimbus.esam.nwu.edu}}[A M Rucklidge and M Silber]

\address{\dag\  Department of Applied
                Mathematics and Theoretical Physics,\hfil\break
                University of Cambridge, Cambridge CB3 9EW, UK}

\address{\ddag\ Department of Engineering Sciences and Applied
                Mathematics,\hfil\break
                Northwestern University, Evanston, IL 60208, USA}

\abs
Motivated by recent analytical and numerical work on two- and
three-dimensional convection with imposed spatial periodicity, we
analyse three examples of bifurcations from a continuous group orbit
of spatio-temporally symmetric periodic solutions of partial
differential equations. Our approach is based on centre manifold
reduction for maps, and is in the spirit of earlier work by Iooss
(1986) on bifurcations of group orbits of spatially symmetric
equilibria.  Two examples, two-dimensional pulsating waves (PW) and
three-dimensional alternating pulsating waves (APW), have discrete
spatio-temporal symmetries characterized by the cyclic groups $Z_n$,
$n=2$ (PW) and $n=4$ (APW).  These symmetries force the Poincar\'e
return map $\bM$ to be the $n^{th}$ iterate of a map~$\bGt$:
$\bM=\bGt^n$.  The group orbits of PW and APW are generated by
translations in the horizontal directions and correspond to a circle
and a two-torus, respectively. An instability of pulsating waves can
lead to solutions that drift along the group orbit, while
bifurcations with Floquet multiplier~$+1$ of alternating pulsating
waves do not lead to drifting solutions.  The third example we
consider, alternating rolls, has the spatio-temporal symmetry of
alternating pulsating waves as well as being invariant under
reflections in two vertical planes. When the bifurcation breaks these
reflections, the map $\bGt$ has a ``two-symmetry,'' as analysed by Lamb
(1996). This leads to a doubling of the marginal Floquet multiplier
and the possibility of bifurcation to two distinct types of drifting
solutions.
\endabs

\submitted

\noindent \today

\AMRsection{Introduction}
Techniques for analysing symmetry-breaking bifurcations of $\Gamma$-invariant
equilibria of $\Gamma$-equivariant differential equations are well-developed in
the case of compact Lie groups~$\Gamma$ (Golubitsky \etal 1988). The motivation
for developing these methods comes, in large part, from problems of pattern
formation in fluid dynamics (see, for example, Crawford \and Knobloch 1991).
In the simplest cases, the symmetry-breaking bifurcation corresponds to a
pattern-forming instability of a basic state that is both time-independent and
fully symmetric, for example, a spatially uniform equilibrium solution of the
governing equations.
A symmetry-breaking Hopf bifurcation of this spatially uniform state often
leads to time-periodic solutions that break the translation invariance of the
governing equations and that have spatio-temporal and spatial symmetries. In
this paper we address bifurcations of such periodic orbits, which have broken
the translation invariance but have retained a discrete group of
spatio-temporal symmetries.
 \addref{refG59} \addref{refC50}

We consider problems posed with periodic boundary conditions, for
which there is an $S^1$ symmetry associated with each direction of
imposed periodicity. If this symmetry is broken by an equilibrium
solution, then the solution is not isolated; there is a continuous
family of equilibria related through the translations.  An instability
of this solution can excite the neutral translations modes(s) and
lead to new solutions that drift along the translation group orbit.
This is the case, for example, in the ``parity-breaking
bifurcation'': a reflection-symmetric steady state undergoes a
symmetry-breaking bifurcation to a uniformly translating
solution. Another example of a bifurcation leading to drift has been
observed in two-dimensional convection: when the vertical mirror plane
of symmetry that separates steady counter-rotating rolls is broken in
a Hopf bifurcation, the resulting solution, called a
direction-reversing travelling wave or pulsating wave (PW), drifts to
and fro (Landsberg \and Knobloch 1991; Matthews \etal 1993). This
periodic orbit is invariant under the combination of advance of half
the period in time with a reflection; any drift in one direction in the first
half of the oscillation is exactly balanced by a drift in the other
direction in the second half, so there is no net drift during the
oscillation.  Similarly in thee-dimensional convection with spatial
periodicity imposed, for example, on a square lattice, a
symmetry-breaking Hopf bifurcation from steady convection in a square
pattern can lead to alternating pulsating waves (APW), which are
invariant under the combination of advance of one quarter the period
and rotation by~$90^\circ$ (Rucklidge 1997). These solutions drift
alternately along the two horizontal coordinate directions, but again
have no net drift over the whole period of the oscillation.
 \addref{refL23} \addref{refM48} \addref{refR37}

There have been a number of studies of bifurcations of compact
group orbits of (relative) equilibria. Iooss (1986) developed an
approach based on centre manifold reduction to investigate
bifurcations of Taylor vortices in the Taylor--Couette
problem. Specifically, he analysed bifurcations in directions
orthogonal to the tangent space to the group orbit of equilibria, with
the neutral translation mode incorporated explicitly in the
bifurcation problem.  Krupa (1990) provided a general setting for
investigating bifurcations of relative equilibria that focuses on the
local dynamics in directions orthogonal to the tangent space to the
group orbit. He shows that the resulting bifurcation problem is
$\Sigma$-equivariant, where $\Sigma$ is the isotropy subgroup of
symmetries of the relative equilibrium, and, building on work of Field
(1980), provides a group theoretic method for determining whether or
not the bifurcating solutions drift. Aston \etal (1992), and Amdjadi
\etal (1997) develop a technique for numerically investigating
bifurcations of relative equilibria in $O(2)$-equivariant partial
differential equations, and apply their method to the
Kuramoto--Sivashinsky equation. Their approach isolates one solution on
a group orbit, while still keeping track of any constant drift along
the group orbit.
 \addref{refI3} \addref{refK61} \addref{refF33}
 \addref{refA40} \addref{refA36}

In this paper we investigate bifurcations of time-periodic solutions 
that are not isolated as they have broken the translation invariance, but that
do possess a discrete group of spatio-temporal symmetries.  Our
approach is similar to that of Iooss (1986). However, we are
interested in instabilities of periodic solutions, so we use centre
manifold reduction for Poincar\'e maps.  We are particularly
interested in determining whether the symmetries of the basic state
place any restrictions on the types of bifurcations that occur, and
whether the bifurcating solutions drift along the underlying group
orbit or not. We consider three examples that are motivated by
numerical studies of convection with periodic boundary conditions in
the horizontal direction(s).  First we investigate bifurcations of the
pulsating waves and alternating pulsating waves described above. These
solutions have discrete spatio-temporal symmetries $Z_2$ and $Z_4$,
respectively.  The group orbit of the pulsating waves is $S^1$, while
the group orbit of the alternating pulsating waves is a two-torus, due
to imposed periodicity in two horizontal directions.  The third
example we treat in this paper is alternating rolls (AR), which have
the same spatio-temporal symmetry as APW but are also invariant under
reflection in two orthogonal vertical planes (Silber \and Knobloch
1991).
 \addref{refS40}

The $Z_n$ ($n=2,4$) spatio-temporal symmetry of the basic state places
restrictions on the Poincar\'e return map $\bM$; specifically, we show
that it is the $n^{th}$ iterate of a map $\bGt$. A direct consequence
of this is that period-doubling bifurcations are nongeneric (Swift
\and Wiesenfeld 1984).  Throughout the paper we restrict our analysis
to bifurcation with Floquet multiplier~$+1$; we do not consider Hopf
bifurcations. We also restrict attention to bifurcations that preserve
the spatial-periodicity of the basic state.
 \addref{refI3} \addref{refS20}

Our paper is organized as follows. In the next section we lay the
framework for our analysis in the setting of a simple example, namely
bifurcation of pulsating waves. We show how the spatio-temporal
symmetry is manifest in the Poincar\'e return map. Section 3 considers
bifurcation of the three-dimensional analogue of pulsating waves,
namely alternating pulsating waves.  Section 4 considers bifurcations
of alternating rolls.  For this problem we need to consider six
different cases, which we classify by the degree to which the spatial,
and spatio-temporal symmetries are broken. In the case that the
spatial reflection symmetries are fully broken by the neutral modes,
the Floquet multiplier~$+1$ is forced to have multiplicity two, and more
than one solution branch bifurcates from the basic AR state. In one
case we find a bifurcation of the AR state leading to two distinct
drifting solutions. We present an example of one of the drifting
patterns that is obtained by numerically integrating the equations of
three-dimensional compressible magnetoconvection.  In the course of
the analysis of bifurcations of alternating rolls, we make contact
with the work on $k$-symmetries of Lamb \and Quispel (1994) and Lamb
(1996).  Section 5 contains a summary and indicates some directions
for future work.
 \addref{refL43}
 \addref{refL41}

\AMRsection{Two dimensions: pulsating waves}
We write the partial differential equations (PDEs) for two-dimensional
convection symbolically as:
 $$
 {\rmd\bU\over\rmd t} = \bF(\bU;\mu),\eqno(\AMReqno)\AMReqlabel{thePDE}
 $$
where $\bU$ represents velocity, temperature, density, etc.\
as functions of the
horizontal coordinate~$x$, the vertical coordinate~$z$ and time~$t$; $\mu$
represents a parameter of the problem; and  $\bF$ is a nonlinear operator
between suitably chosen function spaces. We assume periodic boundary
conditions, with spatial period $\ell$, in the $x$-direction.

The symmetry group of the problem is $O(2)$, which is the semi-direct product
of $Z_2$, generated by a reflection~$\kappa_x$, and an $SO(2)$ group of
translations~$\tau_a$, which act as
 $$
 \kappa_x\colon x\to-x,\qquad
 \tau_a\colon x\to x+a\pmod\ell,\eqno(\AMReqno)
 $$
where $\tau_\ell$ is the identity and $\tau_a\kappa_x=\kappa_x\tau_{-a}$.
The PDEs~\thePDE\ are equivariant under the action of these
symmetry operators, so $\bF(\tau_a\bU;\mu)=\tau_a\bF(\bU;\mu)$ and
$\bF(\kappa_x\bU;\mu)=\kappa_x\bF(\bU;\mu)$, where $\tau_a$ and $\kappa_x$
act on the functions as follows:
 $$
 \tau_a\bU(x,z,t)\equiv\bU(x-a,z,t),\qquad
 \kappa_x\bU(x,z,t)\equiv M_{\kappa_x}\bU(-x,z,t).\eqno(\AMReqno)
 $$
Here $M_{\kappa_x}$ is a matrix representing~$\kappa_x$; it reverses the sign
of the horizontal component of velocity and leaves all other fields in $\bU$
unchanged.

Suppose that when the parameter $\mu=0$, there is a known pulsating wave
solution~$\bU_0(x,z,t)$ of~\thePDE\ with temporal period~$T$ and spatial
period~$\lambda=\ell/N$, where $N$ specifies the number of PWs that fit into
the periodic box. The symmetries of  $\bU_0$ are summarized as follows:
 $$
              \bU_0(x,z,t)=
 \kappa_x       \bU_0(x,z,t+\half T)=
              \bU_0(x,z,t+T)=
 \tau_\lambda \bU_0(x,z,t).\eqno(\AMReqno)
 $$
There is a continuous group orbit of PWs generated by translations:
$\bU_a=\tau_a \bU_0$. We are interested in bifurcations from this group orbit.
Following the approach developed by Iooss (1986) and Chossat \and Iooss (1994)
for studying instabilities of continuous group orbits of steady solutions, we
expand about the group orbit of periodic solutions as follows:
 $$
 \bU(x,z,t)=\tau_{c(t)}(\bU_0(x,z,t) + \bA(x,z,t)).
      \eqno(\AMReqno)\AMReqlabel{theExpansion}
 $$
Here translation along the group orbit is given by~$\tau_{c(t)}$, where $c$ is
a coordinate parameterizing the group orbit. Small perturbations, orthogonal to
the tangent direction of the group orbit, are specified by $\bA(x,z,t)$. The
expansion~\theExpansion\ is substituted into the PDEs~\thePDE\ and, after
suitable  projection that separates translations along the group orbit from the
evolution of the perturbation orthogonal to it, we obtain equations of the
form (see Chossat \and Iooss (1994)):
 $$
 {\rmd \bA\over\rmd t} = \bG(\bA,\bU_0;\mu),\qquad
 {\rmd c\over\rmd t} = h(\bA,\bU_0;\mu),
   \eqno(\AMReqno)\AMReqlabel{thePDEprojected}
 $$
where $\bG$ and $h$ satisfy $\bG(0,\bU_0;0)=0$ and
$h(0,\bU_0;0)=0$. An important consequence of the translation invariance of the
original PDEs is that $\bG$ and $h$ do not depend on the position $c$ along the
group orbit; the equation for the drift $c$ is decoupled from the
equation for the amplitude of the perturbation $\bA$. Here we find it
convenient to keep track of the explicit time dependence of $\bG$ and $h$,
which enters through their dependence on the basic state $\bU_0$,
by listing $\bU_0$ as one of the arguments of $\bG$ and $h$.
We determine how the
spatio-temporal reflection symmetry of $\bU_0$ is
manifest in the equations for
$c$ and $\bA$ by noting that if $\tau_{c(t)}(\bU_0(x,z,t)+\bA(x,z,t))$ is a
solution of the PDEs~\thePDE\ , then so is
 $$
   \kappa_x\tau_{c(t)}(\bU_0(x,z,t)+\bA(x,z,t))
   =\tau_{-c(t)}(\kappa_x\bU_0(x,z,t)+\kappa_x\bA(x,z,t)).
 \eqno(\AMReqno)
 $$
Hence
 $$\eqalign{
 \bG(\kappa_x\bA,\kappa_x\bU_0;\mu)&=\kappa_x\bG(\bA,\bU_0;\mu),\cr
 h(\kappa_x\bA,\kappa_x\bU_0;\mu)&=       -h(\bA,\bU_0;\mu).}
 \eqno(\AMReqno)
 $$
 \addref{refI3}
 \addref{refC72}

\AMRfigure 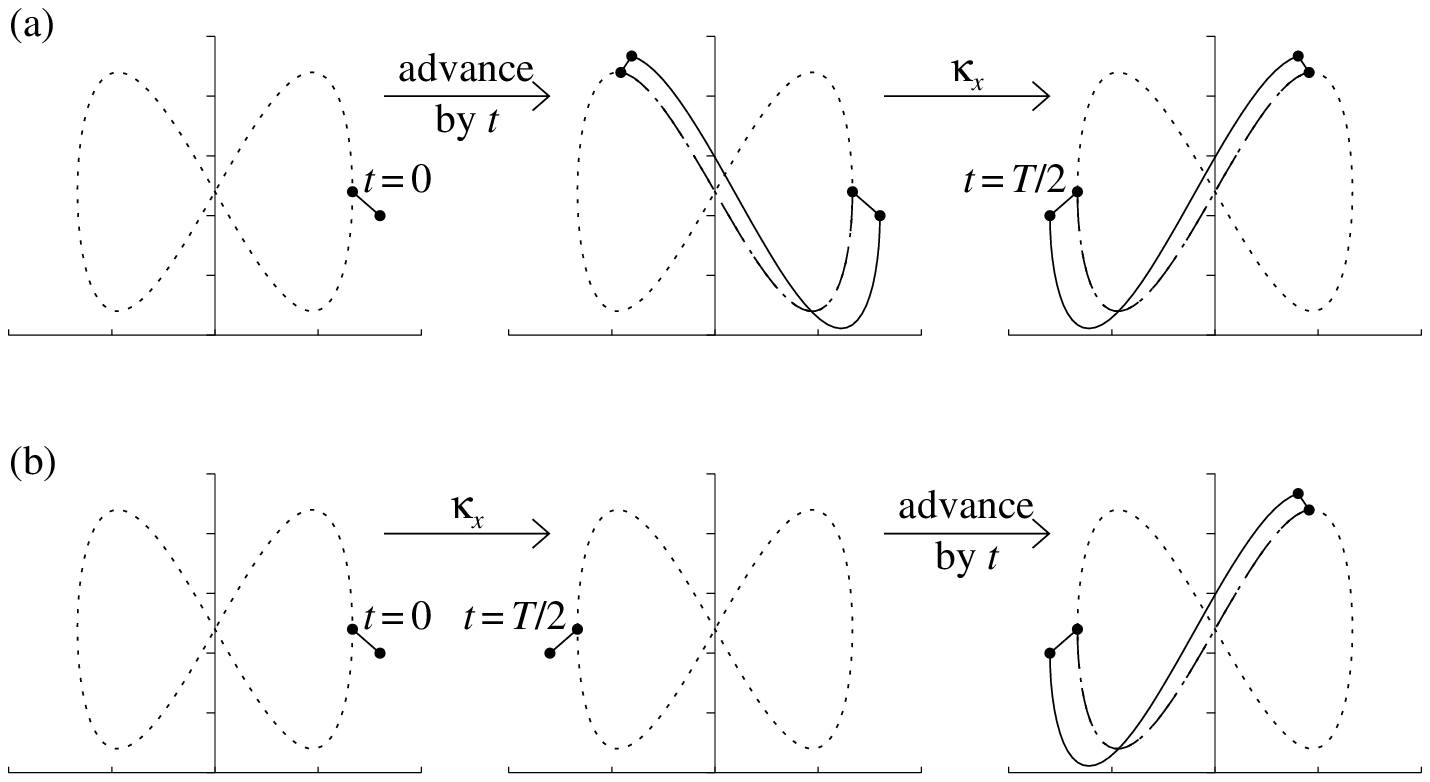
Illustration of
$\kappa_x\bM_0^t = \bM_{T/2}^{T/2+t}\kappa_x.$ In this example,
the reflection~$\kappa_x$ changes the sign of the horizontal coordinate. The
PW periodic solution is shown as a dotted line. (a)~A perturbation at $t=0$ is
advanced in time by an amount~$t$ (the solid line, which stays close to the
broken line on the periodic orbit), then the system is reflected. (b)~We arrive
at the same final position if we reflect (so now the perturbation is about the
PW at $t=\half T$) and then advance in time by the same amount.

\AMRfigurelabel{FigureTimeAdvanceSymmetry}

Since our basic state $\bU_0$ is $T$-periodic, we seek a map that gives the
perturbation $\bA$ at time $t=T$ given a perturbation $\bA(0)$ at some initial
time $t=0$.  Specifically, we define a time advance map $\bM_0^t$ acting on the
perturbation $\bA$ by $\bA(t)=\bM_0^t(\bA(0))$. We adopt the approach of Swift
\and Wiesenfeld (1984) and split the time interval from $0$ to $T$ into two
stages using the symmetry property of the underlying pulsating waves.
Specifically, since $\kappa_x\bA(t)$ satisfies ${\rmd (\kappa_x \bA)\over\rmd
t} =\bG(\kappa_x\bA,\kappa_x\bU_0;\mu)$ and
$\kappa_x\bU_0(x,z,t)=\bU_0(x,z,t+{T\over 2})$, we have $\kappa_x
\bA(t)=\bM_{T/2}^{t+T/2}(\kappa_x\bA(0))$; hence
 $$
 \kappa_x\bM_0^t = \bM_{T/2}^{T/2+t}\kappa_x.\eqno(\AMReqno)
 $$
Advancing the perturbation by a time~$t$ starting from time~$0$ and
then reflecting the whole system is equivalent to reflecting the whole
system then advancing by a time~$t$ starting from time~$\half T$ (see
\FigureTimeAdvanceSymmetry).  It follows immediately that the full
period map $\bM_0^T$ can be written as the second iterate of a map $\bGt$:
 $$
 \bM_0^T=\bM_{T/2}^T\kappa_x^2\bM_0^{T/2}
 =\left(\kappa_x\bM_0^{T/2}\right)^2\equiv\bGt^2.  \eqno(\AMReqno)
 $$
Rather than consider the full period map~$\bM_0^T$, we will
consider the map
$\bGt\equiv\kappa_x\bM_0^{T/2}$. The map~$\bGt$ has no special property under
reflections, but it commutes with translations $\tau_\lambda$, which
leave the underlying pulsating waves invariant:
$\bGt\tau_\lambda=\tau_\lambda\bGt$.
\addref{refS20}

The dynamics of the perturbation is now given by the map~$\bGt$:
$\cA_{n+1}=\bGt(\cA_n;\mu)$, where each iterate corresponds to advancing in
time by $\half T$ and reflecting; thus $\bA(\half T)=\kappa_x\cA_1$,
starting from $\cA_0$ at time~$0$.
In order to
compute the drift~$c_1$ of the solution
at time~$\half T$, we integrate the
$\rmd c/\rmd t$ equation~\thePDEprojected\ for a time
$\half T$, starting at a position~$c_0$ and with initial
perturbation~$\bA(0)=\cA_0$:
 $$
 c_1=c_0 + \int_0^{T/2}h(\bM_0^t(\cA_0),\bU_0(t);\mu)\,\rmd t
    \equiv c_0 + \htid(\cA_0;\mu).\eqno(\AMReqno)
 $$
Then, after a second half-period,
 $$\eqalignno{
 c_2&=c_1 + \int_{T/2}^T h(\bM_{T/2}^t(\bA(\half T)),\bU_0(t);\mu)\,\rmd t\cr
    &=c_1 + \int_{T/2}^T h(\bM_{T/2}^t(\kappa_x\cA_1),
                          \bU_0(t);\mu)\,\rmd t\cr
    &=c_1 + \int_{T/2}^T h(\kappa_x\bM_0^{t-T/2}(\cA_1),
                           \kappa_x\bU_0(t-T/2);\mu)\,\rmd t\cr
    &=c_1 - \int_0^{T/2} h(\bM_0^{t'}(\cA_1),\bU_0(t');\mu)\,\rmd t'\cr
    &=c_1 - \htid(\cA_1;\mu).&(\AMReqno)\cr}
 $$
Thus the combined dynamics of the perturbation and translation can be written
as
 $$
 \cA_{n+1}=\bGt(\cA_n;\mu),\qquad
   c_{n+1}=c_n + (-1)^n\htid(\cA_n;\mu).
 \eqno(\AMReqno)\AMReqlabel{theCombinedMap}
 $$
Since the unperturbed PW is a non-drifting solution of the problem at $\mu=0$
we have $\bGt(0;0)=0$ and $\htid(0;0)=0$. Moreover, the spatial
periodicity
of $\bU_0$ places some symmetry restrictions on $\bGt$ and $\htid$;
specifically,
$\bGt(\tau_\lambda\cA; \mu)=\tau_\lambda\bGt(\cA;\mu)$ and
$\htid(\tau_\lambda\cA;\mu)=\htid(\cA;\mu)$.

We turn now to the codimension-one bifurcations of the PW, which are the
trivial fixed points $\cA=0$, $c=c_0$ of \theCombinedMap\ when $\mu=0$.
The map~\theCombinedMap\
always has one Floquet multiplier (FM) equal to one because of the translation
invariance of the $c$~part of the map. Bifurcations occur when a FM of the
linearization of $\bGt$ crosses the unit circle: either a
$\hbox{FM}=1$, or a $\hbox{FM}=-1$, or there is a pair of complex conjugate FMs
with unit modulus.
Because we have assumed periodic boundary conditions in the original
PDEs, we expect the spectrum of the linearization to be discrete and
the centre manifold theorem for maps to apply. (See Chossat \and Iooss
(1994) for a discussion of the centre manifold reduction in the
similar problem of bifurcations from Taylor vortices.) Let $\zeta$ be
the eigenfunction associated with the critical
\hbox{FM}, so that on the centre manifold, we can write
 $$ \cA_n = a_n\zeta +
    \Phi(a_n),\eqno(\AMReqno)
 $$
where $\Phi$ is the graph of the centre manifold. The unfolded dynamics takes
the form
 $$
 a_{n+1}=\ghat(a_n;\mu),\qquad c_{n+1}=c_n +
 (-1)^n\hhat(a_n;\mu),
 \eqno(\AMReqno)\AMReqlabel{theCMMap}
 $$
where $\ghat$ and $\hhat$ are the maps $\bGt$ and $\htid$ reduced to the centre
manifold; $\ghat$ and $\hhat$ share the same symmetry properties as $\bGt$
and~$\htid$.
 \addref{refC72}

In this paper, we only consider the case where $\tau_\lambda$ acts
trivially. We therefore expect only
generic bifurcations in the map $\ghat$: saddle-node when
$\hbox{FM}=1$, period-doubling when $\hbox{FM}=-1$ and Hopf when there
are a pair of complex FMs. The FMs for the full period map $\bM_0^T$,
which are the squares of the FMs of $\ghat$, will generically be
either one or come in complex conjugate pairs. In particular, we do
not expect $\bM_0^T$ to have a $\hbox{FM}=-1$; this mechanism for
suppressing period-doubling bifurcations was discussed by Swift \and
Wiesenfeld (1984).
 \addref{refS20}

Here we consider  only the cases where $\ghat$ has a
$\hbox{FM}=+1$ or~$-1$. The normal form in the case $\hbox{FM}=1$ is:
 $$
 a_{n+1}=\mu + a_n - a_n^2,\qquad c_{n+1}=c_n +
        (-1)^n\hhat(a_n;\mu),\eqno(\AMReqno)
 $$
to within a rescaling and a
change of sign. The parameter $\mu$ is zero at the bifurcation point,
and the fixed points of the $a$~part of the map are $a=\pm\sqrt{\mu}$
when $\mu$~is positive. The spatial translations are
 $$ c_0,\quad c_1=c_0 + \hhat(a;\mu),\quad c_2=c_0,\quad
        \dots\eqno(\AMReqno)
 $$
We therefore have a $c_0$-parameterized family of solutions that
vanish, in pairs, as $\mu$ is decreased through $\mu=0$.
We interpret this bifurcation by considering the solutions with
$c_0=-{1\over 2}\hhat(a;\mu)$, $a=\pm\sqrt{\mu}$. In this case,
we have a pair of pulsating wave solutions, translated
with respect to the original PW by $\half\hhat(\pm\sqrt{\mu};\mu)$,
which collide in a  saddle-node bifurcation at
$\mu=0$. The remainder of the family of solutions is obtained by
translating this pair.

The case $\hbox{FM}=-1$ is more interesting. The normal form in the
supercritical case is
 $$
 a_{n+1}=(-1+\mu)a_n - a_n^3,\qquad
   c_{n+1}=c_n + (-1)^n\hhat(a_n;\mu),\eqno(\AMReqno)
 $$
with a fixed point $a=0$ and a period-two orbit $a_n=(-1)^n\sqrt{\mu}$. The
dynamics of the spatial translations are
 $$\eqalign{
  c_0,\quad
 &c_1=c_0 + \hhat(a_0;\mu),\quad
  c_2=c_0 + \hhat(a_0;\mu) - \hhat(-a_0;\mu),\quad\cr
 &c_3=c_0 + 2\hhat(a_0;\mu) - \hhat(-a_0;\mu),\quad
 \dots\cr}\eqno(\AMReqno)
 $$
Since $\hhat(0;0)=0$, and generically
${\partial \hhat\over\partial a}(0;0)\ne 0$,
$\hhat(a_0;\mu)$ and $\hhat(-a_0;\mu)$ have opposite sign for small $\mu$;
this represents a symmetry-breaking bifurcation that leads to a
solution that drifts along the group orbit of the PW.

The main points of interest in this section are the approach that we have taken
in analysing the instabilities of the group orbit of the spatio-temporally
symmetric periodic orbit, and the observation that an instability of the
pulsating wave with $\hbox{FM}=1$ in the full-period map can lead to drifting
solutions or not.  Whether solutions drift can only be determined by examining
the half-period map. In the next two sections, we apply our method to
three-dimensional alternating pulsating waves and to alternating rolls, the
latter having spatial as well as spatio-temporal symmetries.

\AMRsection{Three dimensions: alternating pulsating waves}
Alternating pulsating waves (APW) are the simplest three-dimensional
analogue of the pulsating waves discussed in the previous
section. These periodic oscillations have been observed in numerical
simulations of three-dimensional compressible magnetoconvection with
periodic boundary conditions in the two horizontal directions
(Matthews \etal 1995). They appear either after a series of global
bifurcations (Rucklidge \and Matthews 1995; Matthews \etal 1996) or in a Hopf bifurcation
from convection in a square pattern (Rucklidge 1997), and are
invariant under the combined operation of advancing one quarter period
in time and rotating $90^\circ$ in space.
 \addref{refM42} \addref{refR29} \addref{refR37} \addref{refM66}

The full symmetry group of
the problem is the semi-direct product of the $D_4$~symmetry group of
the square lattice and a two-torus~$T^2$ of translations in the two
horizontal directions, $x$ and~$y$. $D_4$ is generated by a
reflection~$\kappa_x$, a clockwise rotation by $90^\circ$~$\rho$,
 $$\eqalign{ &\kappa_x\colon
 (x,y)\to(-x,y),\qquad \rho\colon (x,y)\to(y,-x),\cr &\tau_{a,b}\colon
 (x,y)\to (x+a\pmod\ell,\ y+b\pmod\ell),}\eqno(\AMReqno)
 $$
where $\rho\tau_{a,b}=\tau_{b,-a}\rho$.

As before, we assume that at $\mu=0$, we have a known APW solution
$\bU_{0,0}(x,y,z,t)$ with spatial period~$\lambda$ in each direction and
temporal period~$T$; then $\bU_{0,0}$ satisfies:
 $$\eqalign{
                  \bU_{0,0}(x,y,z,t)&=
 \rho             \bU_{0,0}(x,y,z,t+\quarter T)=
                  \bU_{0,0}(x,y,z,t+T)\cr
 &=\tau_{\lambda,0} \bU_{0,0}(x,y,z,t)=
   \tau_{0,\lambda} \bU_{0,0}(x,y,z,t).}\eqno(\AMReqno)
 $$
There is a two-parameter continuous group orbit of APWs generated by
translations: $\bU_{a,b}=\tau_{a,b}\bU_{0,0}$. We expand about this group
orbit:
 $$
 \bU(x,y,z,t)=\tau_{c_x(t),c_y(t)}(\bU_{0,0}(x,y,z,t) + \bA(x,y,z,t)),
      \eqno(\AMReqno)\AMReqlabel{theExpansionThreeD}
 $$
where $(c_x,c_y)$ is a time-dependent translation around the group orbit and
$\bA$ is the perturbation orthogonal to the tangent plane to the group
orbit.  As before, we separate the evolution of the translations from
that of  the perturbation:
 $$
 {\rmd \bA\over\rmd t} = \bG(\bA,\bU_{0,0};\mu),\quad
 {\rmd c_x\over\rmd t} = h_x(\bA,\bU_{0,0};\mu),\quad
 {\rmd c_y\over\rmd t} = h_y(\bA,\bU_{0,0};\mu),
   \eqno(\AMReqno)\AMReqlabel{thePDEprojectedThreeD}
 $$
where we keep track of the explicit time-dependence of
$\bG$, $h_x$, and $h_y$ through the argument $\bU_{0,0}$.
The spatio-temporal symmetry of the basic state $\bU_{0,0}$ is
manifest in $\bG$, $h_x$ and $h_y$ as follows:
 $$\eqalign{
   \bG(\rho\bA,\rho\bU_{0,0};\mu)&=\rho\bG(\bA,\bU_{0,0};\mu),\cr
   h_x(\rho\bA,\rho\bU_{0,0};\mu)&=  h_y(\bA,\bU_{0,0};\mu),\cr
   h_y(\rho\bA,\rho\bU_{0,0};\mu)&= -h_x(\bA,\bU_{0,0};\mu),\cr
 }\eqno(\AMReqno)
 $$
where $\rho\bU_{0,0}(t+{1\over 4}T)=\bU_{0,0}(t)$.
It is convenient to introduce a complex translation $c\equiv c_x+\i c_y$ and a
corresponding $h\equiv h_x+\i h_y$, so $\rho\tau_c=\tau_{-\i c}\rho$.

As before, we define a time advance map acting on the perturbation so
$\bA(t)=\bM_0^t(\bA(0))$; this has the property
 $$
 \bM_0^t\rho  =\rho  \bM_{T/4}^{T/4+t},\qquad
 \bM_0^t\rho^2=\rho^2\bM_{T/2}^{T/2+t},\qquad
 \bM_0^t\rho^3=\rho^3\bM_{3T/4}^{3T/4+t}\eqno(\AMReqno)
 $$
because of the underlying spatio-temporal symmetry of the \hbox{APW}. The full
period map $\bM_0^T$ is then the fourth iterate of a map $\bGt$:

 $$
 \bM_0^T=\rho^4\bM_{3T/4}^T\bM_0^{3T/4}=\rho\bM_0^{T/4}\rho^3\bM_0^{3T/4}
 =\left(\rho\bM_0^{T/4}\right)^4\equiv\bGt^4.
 \eqno(\AMReqno)\AMReqlabel{thebGtdefinition}
 $$
Instead of $\bM_0^T$, we consider $\bGt\equiv\rho\bM_0^{T/4}$, which
has no special properties under reflections and rotations, but does
commute with $\tau_{\lambda,0}$ and~$\tau_{0,\lambda}$, which leave
the underlying APW invariant.

The dynamics of the perturbation is given by $\cA_{n+1}=\bGt(\cA_n)$, where
$\bA(\quarter T)=\rho^3\cA_1$, etc. Then
 $$
 c_1=c_0 + \int_0^{T/4}h(\bM_0^t(\cA_0),\bU_{0,0}(t);\mu)\,\rmd t
    \equiv c_0 + \htid(\cA_0;\mu),\eqno(\AMReqno)
 $$
where the map $\htid=\htidx+\i\htidy$ is invariant under
the translations $\tau_{\lambda,0}$ and~$\tau_{0,\lambda}$.
After the next quarter period, we find
 $$\eqalignno{
 c_2&=c_1 + \int_{T/4}^{T/2} h(\bM_{T/4}^t(\bA(\quarter T)),\bU_{0,0}(t);\mu)\,\rmd t\cr
    &=c_1 + \int_{T/4}^{T/2} h(\bM_{T/4}^t(\rho^3\cA_1),\bU_{0,0}(t);\mu)\,\rmd t\cr
    &=c_1 + \int_{T/4}^{T/2} h(\rho^3\bM_0^{t-T/4}(\cA_1),\rho^3\bU_{0,0}(t-T/4);\mu)\,\rmd t\cr
    &=c_1 + \i\htid(\cA_1;\mu).&(\AMReqno)\cr}
 $$
So the combined dynamics of the perturbation and the translation can be written
as
 $$
 \cA_{n+1}=\bGt(\cA_n;\mu),\qquad
   c_{n+1}=c_n + \i^n\htid(\cA_n;\mu),\eqno(\AMReqno)
   \AMReqlabel{theCombinedMapThreeD}
 $$
where $\bGt(0;0)=\htid(0;0)=0$.

We consider the bifurcations of~\theCombinedMapThreeD\ only in the
case where $\tau_{\lambda,0}$ and~$\tau_{0,\lambda}$ act trivially.
Note that, as in the case of pulsating waves, the
generic bifurcations of APW are either steady state ($\hbox{FM}=+1$) or Hopf,
since $\bM_0^T=\bGt^4$. We consider  bifurcations with $\hbox{FM}=+1$ of
$\bM_0^T$  only;
generically, these occur when the linearization of $\bGt$ has
a FM of $+1$ or~$-1$.
Near a bifurcation point we reduce the dynamics onto the centre manifold
 $$
 a_{n+1}=\ghat(a_n;\mu),\qquad
   c_{n+1}=c_n + \i^n\hhat(a_n;\mu).\eqno(\AMReqno)\AMReqlabel{theCMMapThreeD}
 $$

When a $\hbox{FM}=1$, once again we have a saddle-node bifurcation, this time
involving pairs of APWs that are translated relative to each other.
If a FM is $-1$, we have
$a_n=(-1)^n\sqrt{\mu}$, and the spatial translations are:
 $$\vcenter{\openup1\jot\halign{%
 $\hfil#$\quad&$\hfil#$&${}#$\hfil&\quad$\hfil#$&${}#$\hfil\cr
 c_0,&
  c_1&=c_0 + \hhat(a_0;\mu),
 &c_2&=c_0 + \hhat(a_0;\mu) +\i \hhat(-a_0;\mu),\cr
 &c_3&=c_0 + \i \hhat(-a_0;\mu),
 &c_4&=c_0, \quad\dots\cr}}\eqno(\AMReqno)
 $$
This solution has no net drift (unlike in the
two-dimensional problem), but travels back and forth
different amounts in the two horizontal
directions since, generically, $\hhat_x(a_0;\mu)\ne\hhat_y(a_0;\mu)$.
The solution remains invariant under advance of half its period in time
combined with a rotation of~$180^\circ$. To see this, we
construct the solution $\bU(x,y,z,t)$ at $t=0$ and $t=\half T$ using
the solution in the $c_0$-parameterized family that satisfies
$c_0=-c_2$. Specifically, we insert the centre manifold
solution $\bA(0)=\cA_0=a_0\zeta
+\Phi(a_0)$, $\bA(\half T)=\rho^2\cA_2=\rho^2(a_0\zeta+\Phi(a_0))$
in~\theExpansionThreeD. We obtain
 $$\eqalignno{
 \bU(0)&=\tau_{c_0}(\bU_{0,0}(0)+a_0\zeta+\Phi(a_0))\cr
 \bU(\half T)&=\tau_{-c_0}(\bU_{0,0}(\half T)+\rho^2
        a_0\zeta+\rho^2\Phi(a_0))\cr
             &=\tau_{-c_0}\rho^2(\bU_{0,0}(0)+a_0\zeta+\Phi(a_0))\cr
             &=\rho^2\bU(0),&(\AMReqno)\cr}
 $$
where we have suppressed the $(x,y,z)$-dependence of $\bU$, retaining
only its $t$-dependence.

Thus, in the simple case of APW, we cannot get drifting solutions in a
bifurcation with $\hbox{FM}=1$ for the time-$T$ return map. We next consider
the same bifurcation for the more complicated example of alternating
rolls. This solution has the same spatio-temporal symmetry as APW but
has extra spatial reflection symmetries. We shall see that in this
case a particular symmetry-breaking bifurcation leads to two distinct
types of drifting solutions.

\AMRsection{Additional spatial symmetries: alternating rolls}
Alternating rolls (AR) are created in a primary Hopf bifurcation from a
$D_4\AMRsemidirectprod T^2$ invariant trivial solution (Silber \and Knobloch
1991). Like alternating pulsating waves, alternating rolls are invariant under
the spatio-temporal symmetry of advancing one-quarter period in time and
rotating $90^\circ$ in space, but have the additional property of being
invariant under reflections in two orthogonal vertical planes.
Alternating rolls have been observed in
three-dimensional incompressible and compressible magnetoconvection (Clune \and
Knobloch 1994; Matthews \etal 1995).
 \addref{refS40} \addref{refM42} \addref{refC49}

For convenience in this section, we define $\rhot$ to be the combined advance
of one quarter period in time followed by a  $90^\circ$ clockwise
rotation about the line
$(x,y)=(0,0)$. Reflecting in the planes $x=\quarter\lambda$ or
$y=\quarter\lambda$ leaves alternating rolls unchanged at all times, so the
sixteen-element group that leaves AR invariant is generated by $\kxp$, $\kyp$
and $\rhot$, where
 $$\eqalign{
 \kxp&\colon     (x,y,z,t)\to(\half\lambda-x,y,z,t),\cr
 \kyp&\colon     (x,y,z,t)\to(x,\half\lambda-y,z,t),\cr
 \rhot&\colon    (x,y,z,t)\to(y,-x,z,t+\quarter T).\cr}\eqno(\AMReqno)
 $$
The basic AR solution $\bU_{0,0}(x,y,z,t)$ exists at $\mu=0$ and satisfies
 $$\eqalign{
                  \bU_{0,0}(x,y,z,t)&=
 \rho             \bU_{0,0}(x,y,z,t+\quarter T)=
                  \bU_{0,0}(x,y,z,t+T)\cr
 &=\kxp \bU_{0,0}(x,y,z,t) = \kyp \bU_{0,0}(x,y,z,t)\cr
 &=\tau_{\lambda,0} \bU_{0,0}(x,y,z,t)=
   \tau_{0,\lambda} \bU_{0,0}(x,y,z,t).}\eqno(\AMReqno)
 $$
As in section~3, we expand about this basic solution and recover the
map~\theCombinedMapThreeD. The presence of extra reflection symmetries of
the underlying solution manifests itself in the following way:
 $$\vcenter{\openup1\jot\halign{%
 $\hfil#$&${}#$\hfil&\qquad$\hfil#$&${}#$\hfil\cr
 \bGt(\kxp\cA)  &=\kyp\bGt(\cA),  &\bGt(\kyp\cA)&=\kxp\bGt(\cA),\cr
 \htidx(\kxp\cA)&=-\htidx(\cA),   &\htidx(\kyp\cA)&=\htidx(\cA),\cr
 \htidy(\kxp\cA)&=\htidy(\cA),    &\htidy(\kyp\cA)&=-\htidy(\cA).\cr
 }}\eqno(\AMReqno)\AMReqlabel{theARSymmetryAction}
 $$
Note that the rotation in the definition of $\bGt$ \thebGtdefinition\ implies
that reflecting with $\kxp$ then applying $\bGt$ is equivalent to applying
$\bGt$ then reflecting with~$\kyp$, since $\rho\kxp=\kyp\rho$. In the
terminology of Lamb \and Quispel (1994), $\kxp$ and~$\kyp$ are 2-symmetries of
$\bGt$, that is, $\bGt^2(\kxp\cA)=\kxp\bGt^2(\cA)$. In general, $k$-symmetries 
arise when the spatial part of the spatio-temporal symmetry of a time-periodic 
solution does not commute with its purely spatial symmetries (Lamb 1997).
 \addref{refL43} \addref{refL42}

\topinsert
\table{Summary of six types of bifurcations of alternating rolls,
distinguished by the action of $\kxp$ and $\kyp$ on the critical
modes, and by the critical Floquet multipliers of $\bGt$.
Isotropy subgroups (up to conjugacy) of the
bifurcating solution branches are indicated, along with their order;
in cases \Bp\ and \Bm, there are two distinct
solution branches.}
 \line{\hfil\vbox{\hrule\smallskip \halign{ \quad#\hfil\quad&%
 #\hfil\quad&#\hfil\quad&#\hfil\quad
&#\hfil\quad\cr
 Case&Action of $\kxp$, $\kyp$ on&Floquet&Bifurcation&Isotropy \cr
    &marginal              modes&multiplier(s)& (drift or not)
 &subgroup (order)\cr
 \noalign{\smallskip}
 \noalign{\hrule}
 \noalign{\smallskip}
\Ap($+1$)&$\kxp\kyp\zeta=\zeta$&$\hbox{FM}=+1$&Saddle-node&$\langle\kxp,\kyp,\rhot\rangle$\hfill(16)\cr
         &$\kxp\zeta=\kyp\zeta=\zeta$&        &(no    drift)&\cr
 \noalign{\medskip}
\Ap($-1$)&as \Ap($+1$)&$\hbox{FM}=-1$&Symmetry-breaking&$\langle\kxp,\kyp,\rhot^2\rangle$\hfill(8)\cr
         &            &              &(no          drift)&\cr
 \noalign{\medskip}
\Am($+1$)&$\kxp\kyp\zeta=\zeta$&$\hbox{FM}=+1$&Symmetry-breaking&$\langle\kxp\kyp,\rhot\rangle$\hfill(8)\cr
         &$\kxp\zeta=\kyp\zeta=-\zeta$&        &(no         drift)&\cr
 \noalign{\medskip}
\Am($-1$)&as \Am($+1$)&$\hbox{FM}=-1$&Symmetry-breaking&$\langle\kxp\kyp,\kxp\rhot\rangle$\hfill(8)\cr
        &             &              &(no          drift)&\cr
 \noalign{\medskip}
\Bp&$\kxp\kyp\zeta_\pm=-\zeta_\pm$&$\hbox{FM}=\pm1$&Symmetry-breaking&$\langle\kxp,\rhot^2\rangle$\hfill(4)\cr
  &$\zeta_-=\kxp\zeta_+=-\kyp\zeta_+$&             &(no   net    drift)&$\langle\rhot\rangle$\hfill(4)\cr
 \noalign{\medskip}
\Bm&as \Bp&$\hbox{FM}=\pm\i$&Symmetry-breaking&$\langle\kyp\rangle$\hfill(2)\cr
  &     &                 &(drift)&Trivial\hfill(1)\cr
 \noalign{\smallskip}
 }\smallskip\hrule}\hfil}
 \endinsert

\AMRtablelabel{TableInstabilities}

The remainder of this section is devoted to the discussion of the
codimension-one steady bifurcations of this problem. We do not
consider bifurcations that break the spatial periodicity, so
$\tau_{\lambda,0}$ and $\tau_{0,\lambda}$ act trivially, nor do we
consider Hopf bifurcations. The results are summarised in
\TableInstabilities.

We begin by noting that $\bGt(\kxp\kyp\cA)=\kxp\kyp\bGt(\cA)$, so $\kxp\kyp$
commutes with the linearisation~$\bLt$ of~$\bGt$,
whereas $\kxp\bLt=\bLt\kyp$.
The eigenspaces of $\bLt$ are invariant under the reflection
$\kxp\kyp$. We assume the generic situation of one-dimensional
eigenspaces, then each
eigenfunction $\zeta$ must be either even or odd under
the reflection $\kxp\kyp$, i.e.,  $\kxp\kyp\zeta=\zeta$ (case~A)
or $\kxp\kyp\zeta=-\zeta$ (case~B), since $(\kxp\kyp)^2$ is the
identity. In case~A, if
$\zeta=\kxp\kyp\zeta$ is an eigenfunction of $\bLt$ with FM~$s$, then
$\kxp\zeta=\kyp\zeta$ has the same FM:
 $$
 \bLt\kxp\zeta=\kyp\bLt\zeta=s\kyp\zeta=s\kxp\zeta.
 \eqno(\AMReqno)
 $$
Therefore, $\zeta$ and $\kxp\zeta$  are
linearly dependent; moreover, $\kxp^2$ is the identity, so either
$\kxp\zeta=\zeta$ (case~\Ap) or $\kxp\zeta=-\zeta$ (case~\Am). Finally, these
two cases are subdivided according to the value of the critical Floquet
multiplier of $\bLt$,
(either $+1$ or~$-1$) at the bifurcation point.

Case~B is rather different. Here we have $\kxp\zeta=-\kyp\zeta$, so
 $$
 \bLt\kxp\zeta=\kyp\bLt\zeta=s\kyp\zeta=-s\kxp\zeta.
 \eqno(\AMReqno)
 $$
Thus $\kxp\zeta$ has $\hbox{FM}=-s$ and is linearly independent of $\zeta$,
which has $\hbox{FM}=s$. We define $\zeta_+$ to be the eigenfunction of $s$ and
$\zeta_-$ to be the eigenfunction of $-s$, with
$\zeta_-=\kxp\zeta_+=-\kyp\zeta_+$. There are two ways in which two Floquet
multipliers $s$ and $-s$ can cross the unit circle: either at $+1$ and~$-1$
(case~\Bp) or at $+\i$ and $-\i$ (case~\Bm). Note that in the absence
of the reflection symmetries these bifurcations would be
codimension-two; here they occur as generic bifurcations.
Since the FMs of the time-$T$ map $\bM_0^T$ are the fourth power of the
FMs of~$\bGt$, the effect of the symmetry in
case~B is to force a repeated $\hbox{FM}=+1$ in the  map $\bM_0^T$.

In case~A, we write
 $$
 \cA_n = a_n\zeta + \Phi(a_n),
 \eqno(\AMReqno)
 $$
near the bifurcation point, where $\Phi$ is the graph of the centre manifold.
On the centre manifold we have $\cA=\kxp\kyp\cA$, so
 $$
 \htidx(\cA)=\htidx(\kxp\kyp\cA)=-\htidx(\kyp\cA)=-\htidx(\cA)=0,
 \eqno(\AMReqno)
 $$
where we have used~\theARSymmetryAction. Thus in case~A, $\htidx$ and $\htidy$ are
identically zero, and no bifurcation will lead to drift along the group orbit
of alternating rolls.

The reflections $\kxp$ and $\kyp$ act trivially in case~\Ap. A $\hbox{FM}=+1$
leads to a saddle-node bifurcation of alternating rolls.
The normal form in the case  $\hbox{FM}=-1$ gives
$a_n=(-1)^na_0$, from which the bifurcating solution
$\bU(t)$ can be reconstructed. Choosing the initial translation
$c_0$ to be zero, and suppressing
the $(x,y,z)$-dependence of $\bU$, we have
 $$\fl\vcenter{\openup1\jot\halign{%
 $\hfil#$&${}#$\hfil&\qquad$\hfil#$&${}#$\hfil\cr
  \bU(0)               &= \bU_{0,0}(0) + a_0\zeta+\Phi(a_0),
 &\bU(\quarter T)      &= \rho^3(\bU_{0,0}(0) - a_0\zeta+\Phi(-a_0)), \cr
  \bU(\half T)         &= \rho^2(\bU_{0,0}(0) + a_0\zeta+\Phi(a_0)),
 &\bU(\threequarters T)&= \rho  (\bU_{0,0}(0) - a_0\zeta+\Phi(-a_0)). \cr
 }}\eqno(\AMReqno)
 $$
Here it should be recalled that
$\bU_{0,0}(\quarter T)=\rho^3\bU_{0,0}(0)$, and that
on the centre manifold
 $$
 \bA(\quarter T)=\rho^3\cA_1
 =\rho^3(a_1\zeta+\Phi(a_1))=\rho^3(-a_0\zeta+\Phi(-a_0)).
 \eqno(\AMReqno)
 $$
This solution satisfies
 $$
 \bU(t)=\kxp\bU(t)=\kyp\bU(t)=\rho^2\bU(t+\half T),
 \eqno(\AMReqno)
 $$
and thus has the
same symmetries as
``standing cross-rolls'', described by Silber \and Knobloch (1991).
 \addref{refS40}

In case \Am, $\kxp$ and $\kyp$ act nontrivially,
so the behaviour on the centre
manifold is governed by a pitchfork normal form ($a_n=a_0$) when the
$\hbox{FM}=+1$ and by  a
period-doubling normal form ($a_n=(-1)^na_0$) when the $\hbox{FM}=-1$.
At leading order in $a_0$, the bifurcating solutions $\bU(t)$ in the
two cases are
 $$\vcenter{\openup1\jot\halign{%
 $\hfil#$&${}#$\hfil&\qquad$\hfil#$&${}#$\hfil\cr
  \bU(0)               &= \bU_{0,0}(0) + a_0\zeta,
 &\bU(\quarter T)      &= \rho^3(\bU_{0,0}(0) \pm a_0\zeta), \cr
  \bU(\half T)         &= \rho^2(\bU_{0,0}(0)  +  a_0\zeta),
 &\bU(\threequarters T)&= \rho  (\bU_{0,0}(0) \pm a_0\zeta). \cr
 }}\eqno(\AMReqno)
 $$
These solutions are not invariant under $\kxp$ or $\kyp$ (since these change
the sign of~$\zeta$), but are invariant under the product~$\kxp\kyp$. In addition,
$\bU(t)=\rho\bU(t+\quarter T)$ in the case $\hbox{FM}=+1$ and
$\bU(t)=\kxp\rho\bU(t+\quarter T)$ in the case $\hbox{FM}=-1$.

Case~B is more interesting. On the two-dimensional centre manifold, we write
 $$
 \cA_n = (-a_n+b_n)\zeta_+ + (a_n+b_n)\zeta_- + \Phi(a_n,b_n);
 \eqno(\AMReqno)
 $$
the form of this expression is chosen for later convenience. The
map~\theCombinedMapThreeD\ reduces to
 $$\fl
 (a_{n+1},b_{n+1})=\ghat(a_n,b_n;\mu),\qquad
 c_{n+1}=c_n + \i^n(\hhatx(a_n,b_n;\mu)+\i\hhaty(a_n,b_n;\mu)).
 \eqno(\AMReqno)\AMReqlabel{theCMMapCaseB}
 $$
Since $\zeta_-=\kxp\zeta_+=-\kyp\zeta_+$, we have
 $$\eqalign{
 \kxp\cA_n &= ( a_n+b_n)\zeta_+ + (-a_n+b_n)\zeta_- + \kxp\Phi(a_n,b_n),\cr
 \kyp\cA_n &= (-a_n-b_n)\zeta_+ + ( a_n-b_n)\zeta_- + \kyp\Phi(a_n,b_n);\cr
 }\eqno(\AMReqno)
 $$
thus
 $$
\kxp(a_n,b_n)=(-a_n,b_n),\qquad
\kyp(a_n,b_n)=(a_n,-b_n).\eqno(\AMReqno)
$$
 From this and from \theARSymmetryAction, we deduce that on the
centre manifold
$$\eqalign{
\hhatx(a_n,b_n)&=-\hhatx(-a_n,b_n)=\hhatx(a_n,-b_n)\cr
\hhaty(a_n,b_n)&=-\hhaty(a_n,-b_n)=\hhaty(-a_n,b_n),\cr}\eqno(\AMReqno)
\AMReqlabel{symmetryhhatxy}
$$
implying that
$\hhatx(0,b;\mu)=0$ and $\hhaty(a,0;\mu)=0$. Moreover, $\ghat$
inherits
the symmetries \theARSymmetryAction\ of $\bGt$:
 $$
 \kyp\ghat(a_n,b_n)=\ghat(\kxp(a_n,b_n)),\qquad \kxp\ghat
 (a_n,b_n)=
 \ghat(\kyp(a_n,b_n)).\qquad
 \eqno(\AMReqno)
 $$
Thus the linearisation $\Lhat$ of $\ghat$ satisfies
 $$
 \left(\matrix{1&0\cr0&-1\cr}\right)\Lhat
   =\Lhat\left(\matrix{-1&0\cr0&1\cr}\right),
 \eqno(\AMReqno)
 $$
which forces $\Lhat$ to be of the form
 $$
 \Lhat=\left(\matrix{0&\alpha\cr\beta&0\cr}\right),
 \eqno(\AMReqno)
 $$
where $a_n$, $b_n$ can be scaled so that $\alpha=1$. There is a
bifurcation
when $\beta=+1$ or $\beta=-1$, yielding FMs $\pm 1$ (case~\Bp)
or $\pm\i$ (case~\Bm), respectively.

In order to analyse the dynamics near the bifurcation point, we compute the
normal form of the bifurcation problems,
expanding~$\ghat$ as a Taylor series in $a$
and~$b$. The reflection symmetry $\kxp\kyp$ prohibits quadratic terms,
and all but two of
the cubic terms can be removed by near-identity transformations. We thus have
the unfolded normal form, truncated at cubic order, in the two cases \Bp\
and~\Bm:
 $$\eqalign{
 a_{n+1}&=b_n,\cr
 b_{n+1}&=\pm(1+\mu)a_n + P a_n^3 + Q a_n b_n^2,\cr
 c_{n+1}&=c_n + \i^n(\hhatx(a_n,b_n;\mu)+\i\hhaty(a_n,b_n;\mu)),\cr
 }\eqno(\AMReqno)\AMReqlabel{theCMMapCaseBNormalForm}
 $$
where $\mu=0$ at the bifurcation point and $P$ and $Q$ are constants.

Lamb (1996) deduced the $(a,b)$ part of this normal form for a local
bifurcation of a map with a $Z_2\times Z_2$ 2-symmetry group,
appropriate to the case under study here. We have chosen our scalings and
near-identity transformations to match Lamb's notation. Lamb described the
period-one, two and four orbits that are created in the bifurcation at $\mu=0$
and calculated their stability as a function of the constants $P$ and~$Q$; we
will interpret those results in terms of bifurcations from, and drift along,
the group orbit of alternating rolls.
 \addref{refL41}

In case \Bp, there are three types of orbits created.
The first is a period-two
orbit $(a_0,0)\leftrightarrow(0,a_0)$, with $0=\mu+Pa_0^2$. From this
and the symmetries \symmetryhhatxy\ of $\hhat$, we deduce
the drift of the solution at each iterate:
 $$\vcenter{\openup1\jot\halign{%
 $\hfil#$\quad&$\hfil#$&${}#$\hfil&\quad$\hfil#$&${}#$\hfil\cr
 c_0,&
  c_1&=c_0 + \hhatx(a_0,0;\mu),
 &c_2&=c_0 + \hhatx(a_0,0;\mu) - \hhaty(0,a_0;\mu),\cr
 &c_3&=c_0 - \hhaty(0,a_0;\mu),
 &c_4&=c_0, \quad\dots\cr}}\eqno(\AMReqno)
 $$
There is no net drift along the group orbit in this case. Moreover,
the $c_0$-parameterized family of solutions drift to and fro in the
$x$~direction only since $c_n-c_0$ is real.
Consider
$c_0=\half(-\hhatx(a_0,0;\mu) + \hhaty(0,a_0;\mu))=- c_2$,
where
$c_0$ is real and thus corresponds to  a translation in the $x$-direction.
The reconstructed solution $\bU(t)$, at leading order in $a_0$, satisfies
 $$\fl\vcenter{\openup1\jot\halign{%
 $\hfil#$&${}#$\hfil&\qquad$\hfil#$&${}#$\hfil\cr
  \bU(0)               &= \tau_{c_0}(\bU_{0,0}(0)       - a_0\zeta_+ + a_0\zeta_-),
 &\bU(\quarter T)      &= \tau_{c_1}\rho^3(\bU_{0,0}(0) + a_0\zeta_+ + a_0\zeta_-), \cr
  \bU(\half T)         &= \tau_{-c_0}\rho^2(\bU_{0,0}(0) - a_0\zeta_+ + a_0\zeta_-),
 &\bU(\threequarters T)&= \tau_{-c_1}\rho  (\bU_{0,0}(0) + a_0\zeta_+ + a_0\zeta_-), \cr
 }}\eqno(\AMReqno)
 $$
so we have $\bU(t)=\kyp\bU(t)=\rho^2\bU(t+\half T)$. The conjugate orbit,
$(0,a_0)\leftrightarrow(a_0,0)$, has
symmetry $\langle\kxp,\rhot^2\rangle$
and does not drift at all in the $x$~direction.

The second and third types of orbit created in case \Bp\ are a period one orbit
$(a_0,a_0)$ and a period two orbit $(a_0,-a_0)\leftrightarrow(-a_0,a_0)$, with
$0=\mu+(P+Q)a_0^2$ in both cases.  These orbits are mapped to each other by
$\kxp$ or by $\kyp$, so we consider only the orbit with period one. The
translations at each iterate are
 $$\vcenter{\openup1\jot\halign{%
 $\hfil#$\quad&$\hfil#$&${}#$\hfil&\quad$\hfil#$&${}#$\hfil\cr
 c_0,&
  c_1&=c_0 + \hhat(a_0,a_0;\mu),
 &c_2&=c_0 + \hhat(a_0,a_0;\mu) + \i\hhat(a_0,a_0;\mu),\cr
 &c_3&=c_0 + \i\hhat(a_0,a_0;\mu),
 &c_4&=c_0, \quad\dots\cr}}\eqno(\AMReqno)
 $$
This orbit also has no net drift, and by choosing $c_0=-\half(1+\i)
\hhat(a_0,a_0;\mu)$,
we have $c_1=\i c_0$. The reconstructed solution $\bU(t)$, at leading
order in $a_0$, satisfies
 $$\fl\vcenter{\openup1\jot\halign{%
 $\hfil#$&${}#$\hfil&\qquad$\hfil#$&${}#$\hfil\cr
  \bU(0)               &= \tau_{c_0}(\bU_{0,0}(0)       + 2a_0\zeta_-),
 &\bU(\quarter T)      &= \tau_{\i c_0}\rho^3(\bU_{0,0}(0) + 2a_0\zeta_-), \cr
  \bU(\half T)         &= \tau_{-c_0}\rho^2(\bU_{0,0}(0) + 2a_0\zeta_-),
 &\bU(\threequarters T)&= \tau_{-\i c_0}\rho  (\bU_{0,0}(0) + 2a_0\zeta_-), \cr
 }}\eqno(\AMReqno)
 $$
so $\bU(t)=\rho\bU(t+\quarter T)$, and the isotropy subgroup is 
$\langle\rhot\rangle$. This solution has the same
symmetries
as the alternating pulsating waves described in section~3, so 
alternating pulsating waves may be created in a symmetry-breaking
bifurcation of alternating rolls. The period-two orbit 
$(a_0,-a_0)\leftrightarrow(-a_0,a_0)$ has the conjugate isotropy subgroup 
$\langle\kxp\kyp\rhot\rangle$.

Finally, we turn to case~\Bm. Here, there are two types of periodic
orbit created in the bifurcation at $\mu=0$, and in this case they are both of
period four. The first orbit is
$(a_0,0)\rightarrow(0,-a_0)\rightarrow(-a_0,0)\rightarrow(0,a_0)$,
with $0=-\mu+Pa_0^2$ in \theCMMapCaseBNormalForm.
The translations are
 $$\fl\vcenter{\openup1\jot\halign{%
 $\hfil#$\quad&$\hfil#$&${}#$\hfil&\quad$\hfil#$&${}#$\hfil\cr
 c_0,&
  c_1&=c_0 + \hhatx(a_0,0;\mu),
 &c_2&=c_0 + \hhatx(a_0,0;\mu) + \hhaty(0,a_0;\mu),\cr
 &c_3&=c_0 +2\hhatx(a_0,0;\mu) + \hhaty(0,a_0;\mu),
 &c_4&=c_0 +2\hhatx(a_0,0;\mu) +2\hhaty(0,a_0;\mu),
\quad\dots\cr}}\eqno(\AMReqno)
 $$
Note that $c_n-c_0$~is real so there is no drift at all in the
$y$~direction, but there is a systematic drift in the $x$~direction. The
reconstructed solution $\bU(t)$ satisfies
 $$\fl\vcenter{\openup1\jot\halign{%
 $\hfil#$&${}#$\hfil&\qquad$\hfil#$&${}#$\hfil\cr
  \bU(0)               &= \tau_{c_0}(\bU_{0,0}(0)       - a_0\zeta_+ + a_0\zeta_-),
 &\bU(\quarter T)      &= \tau_{c_1}\rho^3(\bU_{0,0}(0) - a_0\zeta_+ - a_0\zeta_-), \cr
  \bU(\half T)         &= \tau_{c_2}\rho^2(\bU_{0,0}(0) + a_0\zeta_+ - a_0\zeta_-),
 &\bU(\threequarters T)&= \tau_{c_3}\rho  (\bU_{0,0}(0) + a_0\zeta_+ + a_0\zeta_-), \cr
 }}\eqno(\AMReqno)\AMReqlabel{firstBminus}
 $$
so we have $\bU(t)=\kyp\bU(t)$. A conjugate orbit, started a quarter period
later, has isotropy subgroup $\langle\kxp\rangle$ and drifts
systematically in the
$y$~direction.

The second type of orbit created in case \Bm\ is
$(a_0,a_0)\rightarrow(a_0,-a_0)\rightarrow(-a_0,-a_0)\rightarrow(-a_0,a_0)$,
with
$0=-\mu+(P+Q)a_0^2$. The translations are
 $$\eqalign{
 c_0,\qquad
 c_1&=c_0 + \hhatx(a_0,a_0;\mu) + \i\hhaty(a_0,a_0;\mu),\cr
 c_2&=c_0 + (1+\i)(\hhatx(a_0,a_0;\mu) + \hhaty(a_0,a_0;\mu)),\cr
 c_3&=c_0 + (2+\i)\hhatx(a_0,a_0;\mu) + (1+2\i)\hhaty(a_0,a_0;\mu),\cr
 c_4&=c_0 + (2+2\i)(\hhatx(a,a;\mu) + \hhaty(a_0,a_0;\mu)), \quad\dots\cr}\eqno(\AMReqno)
 $$
This corresponds to a solution that drifts along the diagonal, with a wobble
from side to side as it goes.
At leading order in $a_0$, the reconstructed solution $\bU(t)$
satisfies
 $$\vcenter{\openup1\jot\halign{%
 $\hfil#$&${}#$\hfil&\qquad$\hfil#$&${}#$\hfil\cr
  \bU(0)               &= \tau_{c_0}(\bU_{0,0}(0)       + 2a_0\zeta_-),
 &\bU(\quarter T)      &= \tau_{c_1}\rho^3(\bU_{0,0}(0) - 2a_0\zeta_+), \cr
  \bU(\half T)         &= \tau_{c_2}\rho^2(\bU_{0,0}(0) - 2a_0\zeta_-),
 &\bU(\threequarters T)&= \tau_{c_3}\rho  (\bU_{0,0}(0) + 2a_0\zeta_+), \cr
 }}\eqno(\AMReqno)\AMReqlabel{secondBminus}
 $$
which has fully broken the spatial and spatio-temporal symmetries of
the underlying alternating rolls solution.

We close our discussion of case~\Bm\ by considering spatio-temporal
symmetries of the drifting solutions that are obtained by moving to an
appropriate travelling frame. These symmetries are not listed in
\TableInstabilities.
We first consider the solution \firstBminus\
that drifts in the $y$ direction; it
has the following spatio-temporal symmetry
 $$
 U(t)=\kxp\rho^2\tau_{c_0-c_2}U(t+\half T).
 \eqno(\AMReqno)\AMReqlabel{driftsymmetryone}
 $$
Next we consider the solution \secondBminus\ that drifts along the
diagonal;
it has spatio-temporal symmetry
 $$
 U(t)=\kyp\rho\tau_{\i c_0^*-c_1}U(t+\quarter T).
 \eqno(\AMReqno)\AMReqlabel{driftsymmetrytwo}
 $$

In summary, we have examined the six different cases in which
alternating rolls undergo a bifurcation with $\hbox{FM}=+1$ in the full period
map. All six
bifurcations preserve the underlying spatial periodicity of the
alternating rolls, but may break the  spatial and spatio-temporal
symmetries. The 2-symmetry present in the B cases forces two Floquet
multipliers to cross the unit circle together, and we find two
branches of bifurcating solutions, with distinct symmetry
properties. It is only in case~\Bm, with Floquet multipliers $\pm\i$
in the map $\bGt$, that the bifurcation leads to systematically
drifting solutions: one solution drifts along a coordinate axis, while
the other drifts along a diagonal.

\AMRfigure 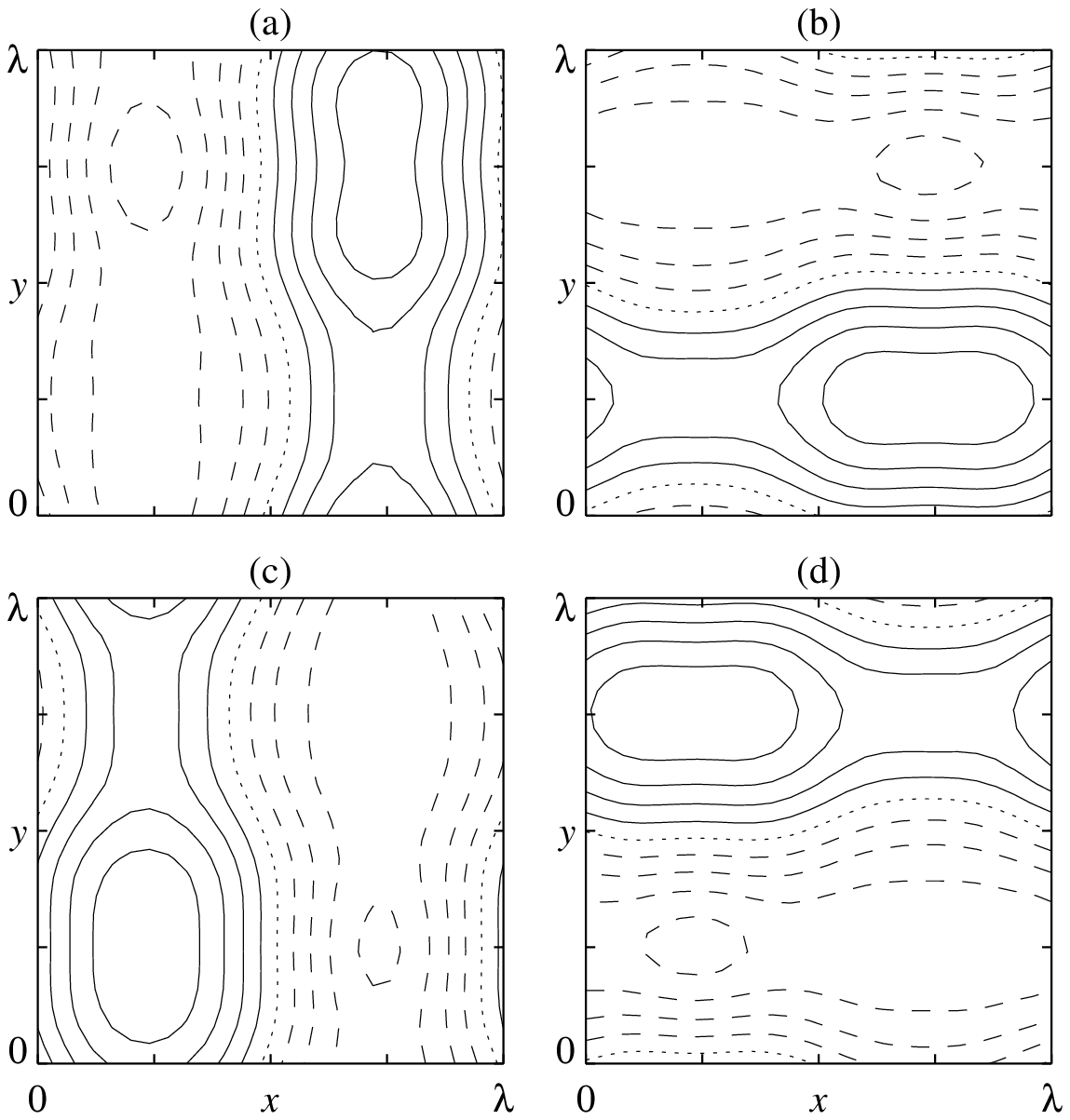
 Alternating rolls in three-dimensional compressible magnetoconvection,
starting with parameter values from Matthews \etal (1995). The four frames are
(approximately) at times (a)~$t=0$, (b)~$t=\quarter T$, (c)~$t=\half T$ and
(d)~$t=\threequarters T$. The frames show contours of the vertical velocity in
a horizontal plane in the middle of the layer: solid lines denote fluid
travelling upwards, dashed lines denote fluid travelling downwards, and the
dotted line denotes zero vertical velocity. The spatial symmetries $\kxp$ and
$\kyp$ are manifest, as is the spatio-temporal symmetry of advancing a quarter
period in time followed by a $90^\circ$ rotation (counter-clockwise in this
example). The dimensionless parameters are: the mid-layer Rayleigh number
(proportional to the temperature difference across the layer) $R=2324$; the
Chandrasekhar number (proportional to the square of the imposed magnetic field)
$Q=1033$; the Prandtl number $\sigma=0.1$; the mid-layer magnetic diffusivity
ratio $\zeta=0.1$; the adiabatic exponent $\gamma=5/3$; the polytropic index
$m=1/4$; the thermal stratification $\theta=6$; the mid-layer plasma beta
$\beta=32$; and the horizontal wavelengths $\lambda=2$ in units of the layer
depth).

\AMRfigurelabel{FigureAR}

\AMRfigure 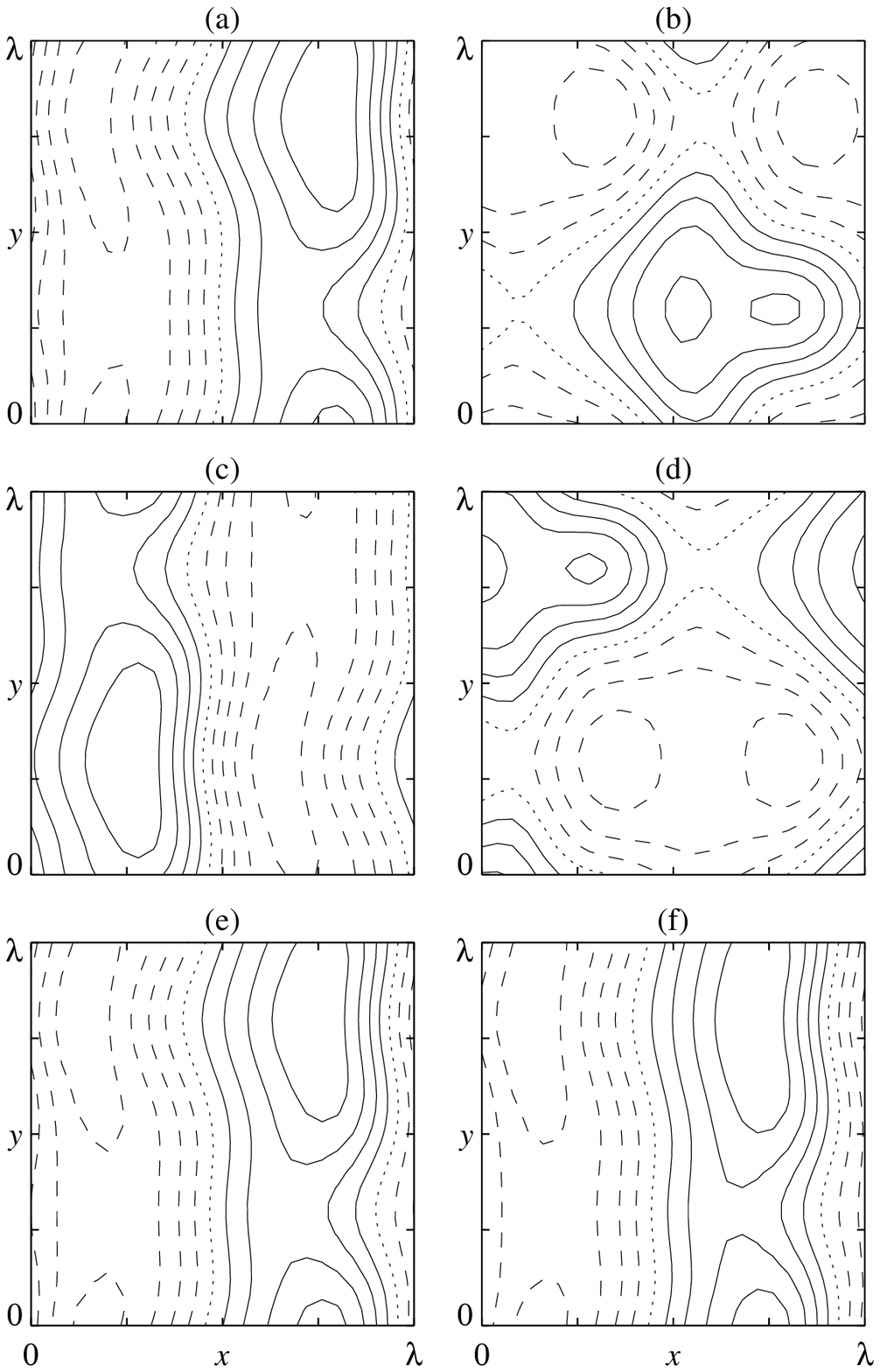
 After a bifurcation of type \Bm, the alternating rolls begin to drift. The
parameter values are as in \FigureAR, but with a higher thermal forcing:
$R=3000$ and $Q=1333$. The frames are (approximately) at times (a)~$t=0$,
(b)~$t=\quarter T$, (c)~$t=\half T$, (d)~$t=\threequarters T$, (e)~$t=T$ and
(f)~$t=2T$. Note how all spatial and spatio-temporal symmetries have been
broken, with the exception of $\kyp$, a reflection in the plane
$y=\quarter\lambda$ (modulo a slight shift in the periodic box). The slow
leftward drift of the pattern can be seen by comparing frames (a), (e) and~(f),
In addition, a drift symmetry $\kxp\rho^2\tau_{c_0-c_2}$, conjugate
to~\driftsymmetryone, can be seen by comparing frames (a) and~(c) or (b)
and~(d).

\AMRfigurelabel{FigureTAR}

We finish this section by presenting an example of a bifurcation from
alternating rolls to a drifting pattern, which we interpret as an instance of a
\Bm\ bifurcation. We have solved the PDEs for three-dimensional compressible
magnetoconvection in a periodic $2\times2\times1$ box, using the code of
Matthews \etal (1995). The PDEs and description of the parameters and numerical
method can be found in that paper. \FigureAR\ shows an example of an
alternating roll at times approximately $0$, $\quarter T$, $\half T$
and~$\threequarters T$; the two reflection symmetries $\kxp$ and $\kyp$ in
planes $x=\quarter\lambda$ and $y=\quarter\lambda$ and the spatio-temporal
symmetry of advancing a quarter period in time followed by a $90^\circ$
rotation about the centre of the box are manifest. Increasing the controlling
parameter, the temperature difference across the layer, leads to the solution
in \FigureTAR: the data are shown at times $0$, $\quarter T$, $\half T$,
$\threequarters T$, $T$ and~$2T$. The only spatial symmetry remaining is the
invariance under~$\kyp$, and the spatio-temporal symmetry has been broken. By
comparing frames (a), (e) and (f) at times $0$, $T$ and~$2T$, it can be seen
that the solution is drifting slowly leftwards along the $x$-axis. Moveover, a
drift symmetry $\kxp\rho^2\tau_{c_0-c_2}$ conjugate to~\driftsymmetryone\ can
be seen by comparing frames (a) and~(c). All evidence points to the bifurcation
being of type \Bm\ (though we have not computed the Floquet multipliers and
critical eigenfunctions in the PDEs).
 \addref{refM42}

\AMRsection{Conclusion}

We have developed a technique for investigating the possible
instabilities from continuous group orbits of spatio-temporally
symmetric time-periodic solutions of partial differential equations in
periodic domains. Our approach is based on centre manifold reduction
and symmetry arguments. It is in the spirit of earlier work by Iooss
(1986) on bifurcations from continuous group orbits of spatially
symmetric steady solutions of partial differential equations.  We have
treated three examples that arise in convection problems: pulsating
waves in two dimensions, and alternating pulsating waves and
alternating rolls in three dimensions. A simple bifurcation can lead
to drifting solutions in the case of pulsating waves but not
alternating pulsating waves. The additional spatial symmetries of
alternating rolls can force two Floquet multipliers to cross the unit
circle together; this degeneracy can lead to drifting solutions, as in
the numerical example presented in the previous section. We have
related our work to the theory of $k$-symmetries developed by Lamb
\and Quispel (1994).
 \addref{refL43}

Our approach can readily be applied to other problems. In the future, we plan
to tackle spatial period doubling and multiplying, where the $\tau_\lambda$
symmetries do not act trivially; such instabilities are relevant to simulations
of convection carried out in larger boxes (Weiss \etal 1996), and will be 
related to the study of the long-wavelength instabilities of alternating rolls
(Hoyle 1994). We also plan to examine the case of the hexagonal lattice: a Hopf
bifurcation on a hexagonal lattice leads to a wide variety of periodic orbits
with different spatio-temporal symmetries (Roberts \etal 1986). Finally, we
plan to investigate the effect of including the extra $Z_2$ mid-layer
reflection symmetry that arises when making the Boussinesq approximation for
incompressible fluids.
 \addref{refW31} \addref{refH26} \addref{refR26}

\ack
It is a great pleasure to acknowledge valuable discussions with
G\'erard Iooss, Jeroen Lamb, and Michael Proctor. We are very grateful to Paul
Matthews for the use of his code for solving the PDEs for three-dimensional
compressible magnetoconvection. This research was
supported by a NATO collaborative research grant CRG-950227. The
research of AMR is supported by the Royal Astronomical Society.  The
research of MS is supported by NSF grant DMS-9404266, and by an NSF
CAREER award DMS-9502266.

\vfill\eject

\references


\def\and{and\ }

\def\JFM{J.~Fluid Mech.}

\def\ARFM{Annu. Rev. Fluid Mech.}

\def\PRS{Proc. R.~Soc. Lond.~A}

\def\PRL{Phys. Rev. Lett.}
\def\PL{Phys. Lett.}
\def\MNRAS{Mon. Not. R. Astron.~Soc.}

\def\AGU{American Geophysical Union}

\def\SIAMJMA{SIAM J.~Math. Anal.}
\def\SIAMJAM{SIAM J.~Appl. Math.}

\newcount\AMRreferenceno

\def\genericref #1#2{\ifundefined{#1}\else%
  \global \advance \AMRreferenceno by 1%
  {#2}\fi}%

\def\author#1#2{#1~#2}%

\outer\def\refarticle#1#2#3#4#5#6#7{%
\genericref{#1}%
{\refjl{#2 #3 #4}{#5}{#6}{#7}}}

\outer\def\refarticleshort#1#2#3#4#5{%
\genericref{#1}%
{\refjl{#2 #3 #4}{#5}{}{}}}%

\outer\def\refbook#1#2#3#4#5#6{%
\genericref{#1}%
{\refbk{#2 #3}{#4}{(#6: #5)}}}%

\outer\def\refartbook#1#2#3#4#5#6#7#8#9{%
\genericref{#1}%
{\refbk{#2 #3 #4}{#5}{ed #6 (#8: #7) pp~\hbox{#9}}}}%

\def\beginrefs{}
\def\inrefsfalse{}
\def\nohyphens{}

\def\i{\ii}

%
%
%

\beginrefs

\begingroup
\nohyphens%
\spaceskip=0.3333em plus 0.25em minus 0.2em
\xspaceskip=0.5em plus 0.15em%

\refarticleshort{refA36}
{\author{F.}{Amdjadi}, \author{P.J.}{Aston} \and
        \author{P.}{Plech\'a$\check{\rm c}$}}{1997}
        {Symmetry breaking Hopf bifurcations in equations with O(2)
        symmetry with applications to the Kuramoto-Sivashinsky
        equation}
        {{\it J.~Comp.~Phys. \rm (in press)}}

\refarticle{refA40}
{\author{P.J.}{Aston}, \author{A.}{Spence} \and \author{W.}{Wu}}{1992}
            {Bifurcation to rotating waves in equations with
                O(2)-symmetry}
            {\SIAMJAM}{52}{792--809}    

\refbook{refC72}
{\author{P.}{Chossat} \and \author{G.}{Iooss}}{1994}
            {The Couette--Taylor Problem}
            {Springer}{New York}

\refarticle{refC49}
{\author{T.}{Clune} \and \author{E.}{Knobloch}}{1994}
            {Pattern selection in three-dimensional magnetoconvection}
            {Physica}{74D}{151--176}

\refarticle{refC50}
{\author{J.D.}{Crawford} \and \author{E.}{Knobloch}}{1991}
            {Symmetry and symmetry-breaking bifurcations in fluid dynamics}
            {\ARFM}{23}{341--387}

\refarticle{refF33}
{\author{M.J.}{Field}}{1980}
            {Equivariant dynamical systems}
            {Trans. Am. Math. Soc.}{259}{185--205}

\refbook{refG59}
{\author{M.}{Golubitsky}, \author{I.}{Stewart} \and \author{D.G.}{Schaeffer}}
    {1988}
    {Singularities and Groups in Bifurcation Theory. Volume~II}
    {Springer}{New York}

\refarticle{refH26}
{\author{R.B.}{Hoyle}}{1994}
            {Phase instabilities of oscillatory standing squares and 
             alternating rolls}
            {Phys. Rev.}{49E}{2875--2880}

\refarticle{refI3}
{\author{G.}{Iooss}}{1986}
            {Secondary instabilities of Taylor vortices into
             wavy inflow or outflow boundaries}
            {\JFM}{173}{273--288}

\refarticle{refK61}
{\author{M.}{Krupa}}{1990}
            {Bifurcations of relative equilibria}
            {\SIAMJMA}{21}{1453--1486}

\refarticle{refL41}
{\author{J.S.W.}{Lamb}}{1996}
            {Local bifurcations in $k$-symmetric dynamical systems}
            {Nonlinearity}{9}{537--557}

\refarticleshort{refL42}
{\author{J.S.W.}{Lamb}}{1997}
            {$k$-Symmetry and return maps of space-time symmetric flows}
            {{\rm (preprint)}}

\refarticle{refL43}
{\author{J.S.W.}{Lamb} \and \author{G.R.W.}{Quispel}}{1994}
            {Reversing $k$-symmetries in dynamical systems}
            {Physica}{73D}{277--304}

\refarticle{refL23}
{\author{A.S.}{Landsberg} \and \author{E.}{Knobloch}}{1991}
            {Direction-reversing traveling waves}
            {\PL}{159A}{17--20}

\refarticle{refM48}
{\author{P.C.}{Matthews}, \author{M.R.E.}{Proctor}, \author{A.M.}{Rucklidge}
 \and \author{N.O.}{Weiss}}{1993}
            {Pulsating waves in nonlinear magnetoconvection}
            {\PL}{183A}{69--75}

\refarticle{refM42}
{\author{P.C.}{Matthews}, \author{M.R.E.}{Proctor} \and
 \author{N.O.}{Weiss}}{1995}
            {Compressible magnetoconvection in three dimensions:
             planforms and nonlinear behaviour}
            {\JFM}{305}{281--305}

\refarticle{refM66}
{\author{P.C.}{Matthews}, \author{A.M.}{Rucklidge}, \author{N.O.}{Weiss}
 \and \author{M.R.E.}{Proctor}}{1996}
            {The three-dimensional development of the shearing instability of
             convection}
            {Phys. Fluids}{8}{1350--1352}

\refarticle{refR26}
{\author{M.}{Roberts}, \author{J.W.}{Swift} \and \author{D.H.}{Wagner}}{1986}
            {The Hopf bifurcation an a hexagonal lattice}
            {Contemp. Math.}{56}{283--318}

\refarticle{refR37}
{\author{A.M.}{Rucklidge}}{1997}
            {Symmetry-breaking instabilities of convection in squares}
            {\PRS}{453}{107--118}

\refartbook{refR29}
{\author{A.M.}{Rucklidge} \and \author{P.C.}{Matthews}}{1995}
            {The shearing instability in magnetoconvection}
            {Double-Diffusive Convection}
            {\author{A.}{Brandt} \and \author{H.J.S.}{Fernando}}
            {\AGU}{Washington}{171--184}

\refarticle{refS40}
{\author{M.}{Silber} \and \author{E.}{Knobloch}}{1991}
            {Hopf bifurcation on square lattices}
            {Nonlinearity}{4}{1063--1106}

\refarticle{refS20}
{\author{J.W.}{Swift} \and \author{K.}{Wiesenfeld}}{1984}
            {Supression of period doubling in symmetric systems}
            {\PRL}{52}{705--708}

\refarticle{refW31}
{\author{N.O.}{Weiss}, \author{D.P.}{Brownjohn},
 \author{P.C.}{Matthews} \and \author{M.R.E.}{Proctor}}{1996}
            {Photospheric convection in strong magnetic fields}
            {\MNRAS}{283}{1153--1164}

\message{These should be the same: \the\AMRreferenceno\space\the\AMRreferencecount}

\endgroup
\inrefsfalse

\bye